\documentclass{ws-p8-50x6-00nofoot}
\begin{document}
\title{SUMMARY OF THE STRUCTURE FUNCTION SESSION AT 
DIS01\footnote{Invited talk presented at the 9th International Workshop on 
Deep Inelastic Scattering (DIS 2001), Bologna, Italy, 27 Apr - 1 May 2001, 
to appear in the Proceedings.}
}
\author{Kenneth Long}
\address{Imperial College London, UK, E-mail: K.Long@ic.ac.uk}
\author{Richard Nisius}
\address{CERN, Switzerland, E-mail: Richard.Nisius@cern.ch}
\author{W. James Stirling}
\address{IPPP Durham, England, E-mail: W.J.Stirling@durham.ac.uk}
\maketitle
\abstracts{The status and ongoing developments in the field of
           deep inelastic scattering presented at the DIS01 workshop
           in Bologna are discussed from both the experimental and the 
           theoretical perspective.
          }
%
%
\vspace{-9cm}\begin{flushright}
{\bf IPPP/01/34} \\ {\bf DCPT/02/68} \\ {\bf IC/HEP/01-2}
\end{flushright}\vspace{6.5cm}
%
%
\section{Introduction}
\label{sec:intro}
 In the last year considerable progress has been made in many
 aspects of the study of the structure functions of photons and 
 nuclei.
 The most important results presented in the structure function 
 session at the DIS01 workshop in Bologna are reviewed.
 The new theoretical developments are discussed in Section~\ref{sec:theory}.
 New experimental information on photon structure is summarised in
 Section~\ref{sec:phst} while Section~\ref{Sect:Prot} reviews the new 
 experimental results pertaining to the structure of the proton.
 The material presented here is intended only as an overview,
 and reflects the personal view of the working group convenors.
 Many of the important
 details of the investigations have of necessity had to be omitted.
 The reader is referred to the write-ups of the individual presentations
 that are to be found elsewhere in these proceedings.
%
%
\section{Theory}
\label{sec:theory}

It is remarkable that a theoretical framework that was established more than twenty
years ago -- short-distance factorisation and next-to-leading (NLO) DGLAP  -- still
gives a very good description of almost all deep inelastic structure function data, for
example  see Section~\ref{Sect:Prot} below. In recent years attention has focused on
calculations that attempt to improve the theoretical prediction by going beyond the
standard `massless quark NLO--DGLAP' framework. Previous DIS Workshops reported the
significant progress on incorporating non-zero quark masses (particularly $m_c$) into
the analysis, and the resulting improved calculations are now used routinely in the
experimental analyses. The outstanding theoretical deficiencies and  uncertainties can
be classified into various types. Within leading-twist perturbation theory, the obvious
goal is a complete NNLO description. At least for massless quarks, this seems to be
within reach (see below) and the likely phenomenological impact has already been
investigated. A variation of this approach is to focus on the kinematic $x \to 0,\; 1$
limits and resum the leading logarithmic corrections to all orders in perturbation
theory, to investigate whether the fit to data is improved. Non-perturbative
corrections arise in various ways, for example higher twist $\left(1/Q^{2}\right)^n, \
n\geq 1$ power corrections, and also heavy nuclear target corrections. A number of
theoretical contributions to this session addressed these issues, and the main results
of these studies are summarised briefly below.

\begin{figure}[htb]
\begin{center}
{\includegraphics[width=0.70\linewidth]{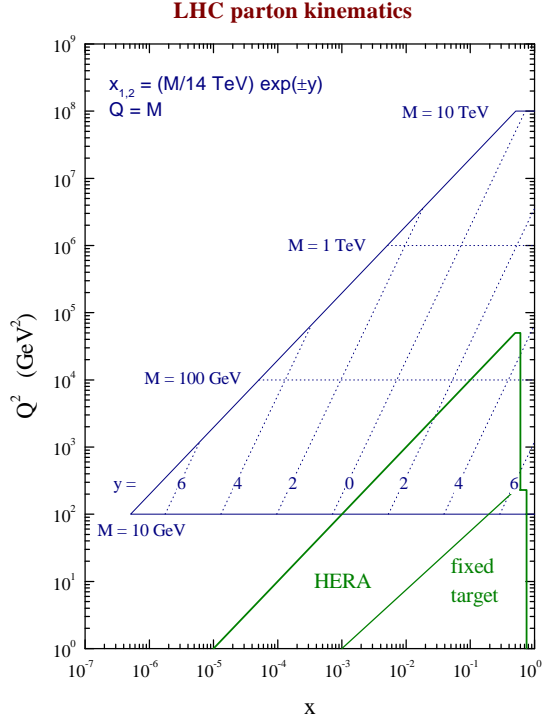}} \caption{ $x$ and $Q^2$ values
probed in the production of an object of mass $M$ and rapidity $y$ at
LHC.\label{fig:js01}}
\end{center}
\end{figure}

Parton distribution functions (PDFs) extracted from deep inelastic scattering structure
functions data play a central role in the calculation of hard scattering cross sections
at the Tevatron and LHC. A precise knowledge of the PDFs is absolutely vital for
reliable predictions for signal and background cross sections. In many cases, it is the
uncertainty in the input PDFs that dominates the theoretical error on the prediction.
Such uncertainties can arise both from the starting distributions themselves,
reflecting the uncertainties in the data input to the global fit, and from evolution to
the higher $Q^2$ scales typical of hadron collider hard scattering processes, which is
sensitive to  uncertainties in $\alpha_s$, unknown higher-order corrections, other
parton flavours, etc. By way of illustration, Figure~\ref{fig:js01} shows the $(x,Q^2)$
values corresponding to the production of heavy objects (e.g. a $W$ or Higgs boson, a $
t\bar t$ pair, a multijet final state etc.) of mass $M$ and rapidity $y$. We assume
leading order kinematics, so that $x = M\exp(\pm y)/\sqrt{s}$ and $Q=M$. As an example,
a $W$ boson ($M = 80$~GeV) produced at rapidity $y=3$ corresponds to the annihilation
of quarks with $x=0.00028$ and $0.11$, probed at $Q^2 = 6400$~GeV$^2$. Notice that in
this example, quarks with these $x$ values are already `measured' in deep inelastic
scattering (at HERA and in fixed--target experiments respectively), but at much lower
$Q^2$.

A rigorous and global treatment of {\it PDF uncertainties} remains elusive, but there
has been a significant advance in the last few years, with several groups now
introducing sophisticated statistical analyses into global or quasi-global fits.
Further progress was reported at this meeting and is summarised below.
%

\subsection{Higher orders, higher twists}
\label{sec:hoht}

In order to match the precision of present and future DIS data, it is necessary to go
beyond NLO in DGLAP perturbation theory. The most relevant quantities here are the NNLO
contribution to the DGLAP splitting functions, i.e. the functions $P^{(2)}$ in the
expansion
\begin{equation}
\mbox{Splitting~Function:} \qquad  P(x,\alpha_s)  = P^{(0)} + \alpha_s P^{(1)}(x) +
\alpha_s^2
 P^{(2)}(x) + ...
\end{equation}
together with the corresponding coefficient functions:
\begin{equation} \mbox{Coefficient~Function:} \qquad  \hat\sigma  = \alpha_s^n\;\left[
\hat\sigma^{(0)} + \alpha_s \hat\sigma^{(1)} + \alpha_s^2  \hat\sigma^{(2)} + ...\;
\right]
\end{equation}
The NNLO contributions to the latter are easier to calculate and are already known for
many of the  cross sections and structure functions of interest.\footnote{Note however
that $ \hat\sigma^{(2)}  $ is still not calculated for some of the key cross sections
used in global fits, e.g.  $d\sigma/dE_T^{jet}$, $d\sigma/dy_W$,
$d^2\sigma^{DY}/dM\,dy$, etc.}

Using recent $N=10,12$ moment calculations by  Retey and  Vermaseren\cite{bib:retey},
van Neerven and Vogt\cite{bib:vogt}  have updated their approximations for
$P^{(2)}(x)$, in which the moment information is combined with sum rule and symmetry
constraints and known leading behaviour for $x\to 0,1$. The resulting approximate
functions are certainly sufficient for $x> 10^{-2}$ phenomenology, and probably
adequate for $ x > 10^{-4}$ as well. Further information can be found in the
contribution by Vogt to these proceedings\cite{bib:vogtp}. It was reported by Vogt at
the meeting that the exact calculation of the NNLO splitting functions should be
completed by the end of the year.

\begin{figure}[htb]
\begin{center}
\vspace{-0.5cm}
 {\includegraphics[width=0.70\linewidth]{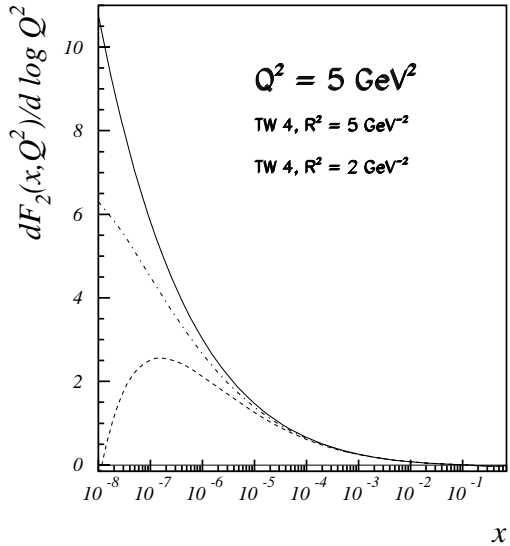}} \caption{Contributions
to the slope
 $\partial F_2/ \partial\log Q^2$ from twist--2 (solid line) and twist--4
 (dash-dotted and dotted lines) operators, with two values of the twist--4 mass scale
 $R^2$ for the latter, from the study by Bl\"umlein
{\it et al.}\protect\cite{bib:blumlein}. \label{fig:js02}}
\end{center}
\end{figure}

Fully {\it quantitative} predictions for higher-twist contributions to DIS structure
functions remain elusive. A potentially important twist--4 contribution comes from the
4--gluon operator matrix element combined with the 4~gluon $\to$ 2~quark coefficient
function. Bl\"umlein\cite{bib:blumleinp} presented an estimate of this contribution to
the structure function slope   $\partial F_2/ \partial\log Q^2$.
 The calculation\cite{bib:blumlein} is performed
using time--ordered perturbation theory and is limited to $dp_{T}^2/p_{T}^4$ accuracy
in the calculation of the corresponding Feynman diagrams. Numerical results are
obtained by modeling the  $4g$ distribution amplitude. At small $x$ it is the {\it
screening} (rather than anti-screening) terms which dominate and reduce the increase of
the slope due to the leading twist--2 contributions. The numerical results, see
Figure~\ref{fig:js02}, show that this effect is significant only below $x \sim 2 \cdot
10^{-6}$ for $Q^2 \sim 5$~GeV$^2$. This is beyond the kinematic range that can be
probed at  HERA but may be accessible at future lepton--hadron  colliders operating at
higher energies.
%

Another type of power correction, of more universal origin, can arise from analytic
resummation of perturbative contributions to all orders. Such resummations typically
lead to  expressions of the form
\begin{equation}
 f(Q^2) \propto  \int_0^{Q^2} \frac{dk^2}{k^2} (k^2)^a \alpha_s(k^2)
\end{equation}
which are ill-defined because of  the Landau pole in the running coupling. Expanding
$\alpha_s(k^2)$ in powers of $\alpha_s(Q^2)$ yields a perturbation series for $f$ whose
coefficients grow as $n!$. The resulting ambiguity in the resummed perturbative result
signals a nonperturbative power correction of the form $(\Lambda^2/Q^2)^a$. An
alternative way to tackle this problem, in the context of QCD quark form factors, using
dimensional regularisation was presented by Magnea\cite{bib:magneap} (see also
\cite{bib:magnea}). If the calculation is performed consistently in $d=4-2\epsilon > 4$
dimensions then the pole moves off the real axis and analytic resummed expressions for
the  form factors can be obtained in terms of gauge and  renormalisation-group
invariant quantities. Applying these techniques to the physically relevant DIS
structure function ($N$) moments yields the standard expression $\propto
N(\Lambda^2/Q^2)$ for the leading power correction at large $N$. The generalisation to
more complicated QCD amplitudes, for example those involving at least two coloured
particles, is more difficult and more work needs to be done.

Small-$x$ structure function phenomenology incorporating higher-twist contributions was
the subject of a presentation by Kotikov\cite{bib:kotikovp}. The starting point of the
analysis is the set of analytic solutions of the DGLAP equations at small $x$ obtained from a
flat parton input at $Q_0^2$~\cite{bib:ballforte}. These solutions are supplemented by
higher-twist (twists 4 and 6) contributions as obtained in a renormalon
model\cite{bib:beneke}. A very reasonable few-parameter fit to HERA structure function
data is obtained (see \cite{bib:kotikovp}), suggesting that higher-twist contributions
do contribute to the observed rise of the $F_2$ structure function at low values of $x$
and $Q^2$.

Another example of QCD resummation was discussed by Ermolaev\cite{bib:ermolaevp} in the
context of the small-$x$ asymptotics of the non-singlet unpolarised and polarised
structure functions. Accounting for both double- and single-logarithmic contributions
and {\it using a fixed QCD coupling} yields the asymptotic behaviour,
\begin{eqnarray}
F_1^{NS}, g_1^{NS} &\sim& \left( \frac{1}{x}\sqrt{\frac{Q^2}{\mu^2}}\right)^{a_\pm}\;
\\
a_+ = \sqrt{2\alpha_s C_F/\pi},&& \ a_- = a_+\; \sqrt{1 + 1/(2N^2)}
\end{eqnarray}
The situation is more complicated when $\alpha_s$ is allowed to run with the internal
ladder transverse momenta as arguments\cite{bib:ermolaev}. Two-dimensional infrared
evolution equations are constructed and solved numerically. The exponents $a_\pm$
become functions of $\Lambda_{\rm QCD}$ and $\mu$, the infrared cut-off parameter.
Quantitatively, $a_{\pm} \to \omega_{\pm}(\mu) = 0.37/0.4$ at 1~GeV, which in fact is
in good agreement with the phenomenological values extracted from fits to $xF_3$ data
at small $x$.

Finally, it is by now well established that it is difficult to disentangle the
underlying small-$x$ QCD dynamics (BFKL, CCFM, DGLAP, etc.) from inclusive quantities
(e.g. $F_2$) alone. More exclusive quantities, for example transverse energy flow and
dijet production rates and correlations, offer a more decisive test, at least in
principle. Szczurek\cite{bib:szczurekp} presented the results of a
study\cite{bib:szczurek} of how dijet azimuthal correlations $d\sigma / d \phi_{\rm
jj}$ in real and virtual photoproduction at HERA can be used to probe the unintegrated
gluon $g(x,k_{\perp}^2)$ in the proton. The benchmark distribution used in the study is
the two-component (soft $+$ hard) model of Ivanov and Nikolaev\cite{bib:nikolaev},
which leads to distinctive kinematical dependence of the dijet correlations. It will be
important to  make comparisons with  the predictions of other models, and of course
with data.

\subsection{PDF fits and uncertainties}
\label{sec:pdfs}
 Progress in obtaining an improved quantitative
understanding of parton distribution functions and their uncertainties was reported at
the meeting. In the context of `best global fits', Thorne\cite{bib:thornep} reported on
a recent update of the MRST (NLO--DGLAP) global fit. The main new ingredients in the
fit are (a) new small-$x$ structure function data from HERA, and  (b) new high-$E_T$
jet data from the Tevatron (see Section~\ref{Sect:AlpSandxG} below for more details).
For the latter, a proper treatment of the correlations between the various systematic
errors is essential and is incorporated in the MRST fitting procedure. The main effect
of the new data is to further constrain the gluon distribution at both small and large
$x$. As for previous global analyses, the overall fit is very good, but there are some
interesting new features: (i) the negativity of the small-$x$ (NLO, $\overline{\rm
MS}$) gluon near the starting scale at 1~GeV$^2$ is now firmly established (although
the gluon becomes positive for all $x > 10^{-5}$ for $Q^2 > 2 - 3$~GeV$^2$); (ii) there
is a slight `tension' between the $\alpha_s(M_Z^2)$ values preferred by the DIS (0.121)
and jet (0.117) data.

\begin{figure}[htb]
\begin{center}
\vspace{1cm}
{\includegraphics[width=0.60\linewidth]{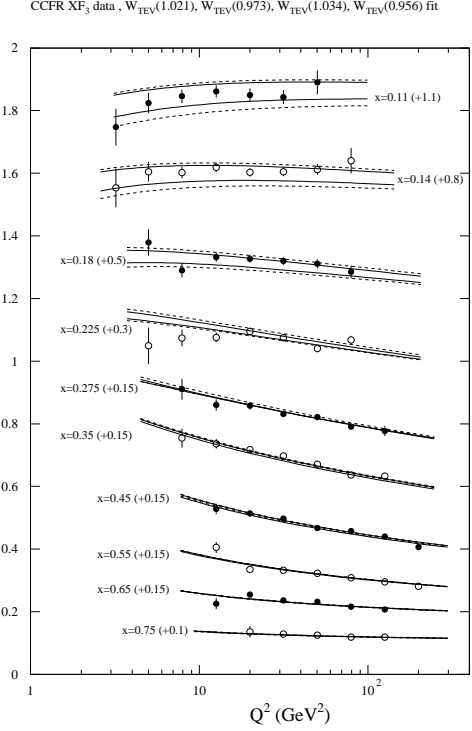}}\vspace{-1cm}\caption{Comparison of
CCFR $F_3(x,Q^2)$ data with theory for $\sigma_W({\rm TEV})$ changing in magnitude by
factors of $1.021$, $0.973$, $1.034$ and $0.956$, from the new MRST study reported by
Thorne\protect\cite{bib:thornep}. \label{fig:js03}}
\end{center}
\end{figure}

Turning to uncertainties on (or due to) PDFs, the MRST approach\cite{bib:MRST2000} (see
also \cite{bib:CTEQb,bib:gieleb})  has been to focus on the uncertainties on particular
physical quantities $\delta \sigma_{\rm obs}$ (for example, the Higgs or weak boson
production cross sections), rather than on uncertainties on the PDFs themselves,
$\delta f_i\;$\cite{bib:botje,bib:alekhin,bib:CTEQa,bib:barone,bib:gielea}. Thorne
reported on a new determination of the PDF uncertainty on $\sigma_W$ at the Tevatron
and LHC. A previous investigation of this uncertainty using the Lagrange Multiplier
method\cite{bib:CTEQb} (see below) had resulted in  $\delta\sigma_{\rm PDF} \simeq 5 -
8 \%$. The MRST approach is to determine which parts of the global fit fail when trying
to force higher/lower $\sigma_{\rm obs}$, and to impose the criterion that no
individual data set has a less than $1\%$ confidence level. This leads to an estimate
of the PDF dependent uncertainty in the total $W$ cross section of $\simeq \pm 2\%\; $
at both the Tevatron and LHC. An example of how one of the components of the MRST
global fit fails when $\sigma_W$ is forced up and down from its central prediction is
shown in Figure~\ref{fig:js03}.
%

\begin{figure}[htb]
\begin{center}
\vspace{1cm}
{\includegraphics[width=0.70\linewidth]{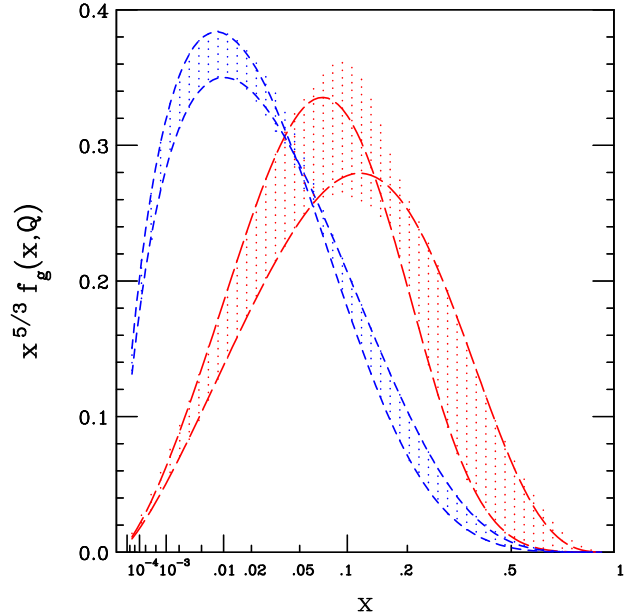}}\vspace{0cm}\caption{Extreme gluon
distributions at $Q=2$ and 10~GeV, from the CTEQ Hessian Matrix error
analysis\protect\cite{bib:CTEQa}. The curves show the extremes at some particular $x$,
while the shaded regions represent the envelope of such extremes.\label{fig:js04}}
\end{center}
\end{figure}

Recent work on parton distribution uncertainties in the context of the CTEQ global fit
was reviewed by Stump\cite{bib:stumpp}. Two methods have been adopted. In the Lagrange
Multiplier constrained fitting approach\cite{bib:CTEQa} the global fit is repeated for
successive  constrained values of a physical observable $\sigma_X$, and a $\chi_{\rm
global}^2$ {\it vs.} $\sigma_X$ profile is established. The issue is then what
`tolerance' (increase in $\chi_{\rm global}^2$ from the minimum) corresponds to a
standard confidence level. In the CTEQ study, the tolerance is determined by
considering the $\chi^2$ response of the individual experiments to the sample PDFs. The
$90\%$ CL allowed ranges for each experiment are then combined into a single overall
tolerance. The second `Hessian Matrix' approach follows a more traditional error
analysis by focusing on the allowed ranges of the individual parameters  $a_1, ... a_d$
that define the PDFs:
\begin{equation}
\chi^2_{\rm global} = \chi_0^2 + \sum_{i,j=1}^{d}(a_i - a_i^0)H_{ij} (a_j - a_j^0) +
\ldots
\end{equation}
Assuming a quadratic approximation, the error on a given observable $X$ is then
\begin{equation}
(\Delta X)^2 = \Delta\chi^2 \sum_{i,j} {\partial X\over \partial a_i} \left(
H^{-1}\right)_{ij} {\partial X\over \partial a_j}\label{eq:hessian}
\end{equation}
The eigenvectors of the Hessian matrix $H$ define orthogonal directions in parameter
space that can be used to define a set of $(2d+1)$ PDF basis sets. Again there is the
issue of the definition of the standard tolerance, i.e. the allowed variation of
$\Delta\chi^2$ in Equation~\ref{eq:hessian}. The PDF basis sets can also be used to
define bands of uncertainty for individual PDFs as the envelope of the extreme curves
at given values of $x$. An example is shown in Figure~\ref{fig:js04}. Notice how the
band shrinks as $Q^2$ increases, and also how the choice of parametric form causes a
slight (artificial) decrease in the band width between the low and high $x$ regions.
%

Finally, in a thought-provoking contribution Collins\cite{bib:collinsp} (see also
\cite{bib:collins}) addressed the issue of how to determine how `good' a good global
fit really is. The point is that a reasonable overall $\chi_{\rm tot}^2$ can mask the
fact that certain subsets of data (i.e. individual experiments) are in strong
disagreement with the theory (which could of course have various explanations, ranging
from errors or deficiencies in the experimental analysis or theoretical calculation to
an unexpected contribution from new physics). The procedure is therefore to
 consider the quality of fit to individual data subsets, explore the parameter space to
 find PDF sets that are
 `best fits' to these individual data sets and to examine by how much the fit to
 a particular data set is improved relative to the increase in overall $\chi_{\rm tot}^2$.
A substantial decrease in an individual $\chi^2$ is a symptom of a bad fit to that
particular data set, which could then be further studied to investigate the source of
the disagreement. It is interesting to note that both the CTEQ and MRST global fits
contain `problem' data subsets according to this definition.

\subsection{Parton distributions in heavy nuclei}
\label{sec:heavy}

%
\begin{figure}[htb]
\begin{center}
{\includegraphics[width=0.70\linewidth]{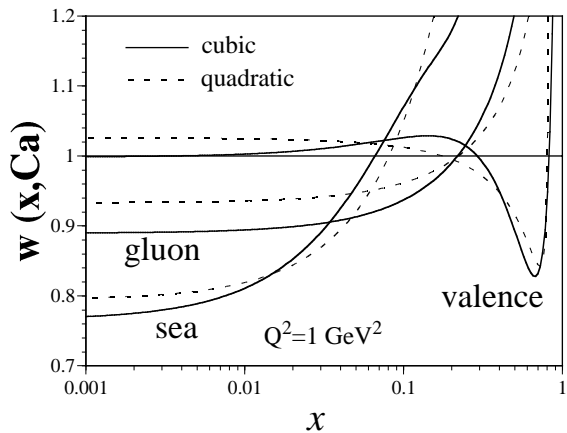}} \caption{The weight functions of
Equation~\ref{eq:weights} for parton distributions in calcium, from the study by Kumano
{\it et al.}\protect\cite{bib:kumano}. \label{fig:js05}}
\end{center}
\end{figure}

When deep inelastic neutrino scattering structure function data are used in a global
PDF fit,  account must be taken of the fact that the PDFs in heavy nuclei are {\it
different} from those in a nucleon. It is important to quantify these differences, not
only for use in global PDF fits but also, for example, when interpreting possible
quark-gluon plasma signals from heavy ion collisions. There have been several recent
analyses to determine the parton distributions in
nuclei\cite{bib:kumanop,bib:kumano,bib:eskola}. Kumano\cite{bib:kumanop} reported the
results of a study\cite{bib:kumano} based on fits to electron and muon $F_2$ data on
nuclear targets. A simple  ansatz is used to parameterise the nuclear modification of
the PDFs:
\begin{eqnarray}
f_i^A(x,Q^2)\; &=& \; w_i(x,A) \; f_i^N(x,Q^2) \nonumber \\
w_i(x,A) \; &= &\; 1 + \left( 1- A^{-1}\right)^{1/3} h_i(x) \label{eq:weights}
\end{eqnarray}
where $h_i$ is taken to be either a quadratic or cubic polynomial in $x$ divided by a
power of $(1-x)$. The parameters in $h_i$ are determined by a $\chi^2$ minimisation
procedure. A reasonable overall fit is obtained ($\chi^2/{\rm dof} = 580/302$), with
the excess $\chi^2$  mainly due to apparent inconsistencies between data sets, e.g.
E665 {\it vs.} NMC. The resulting weight functions for calcium are shown in
Figure~\ref{fig:js05}. As only structure function data are used in the fit, there are
large uncertainties on the extracted gluon and sea-quark weight functions. Comparisons
with Drell-Yan, prompt-$\gamma$ and heavy quark production data are needed to further
constrain $g^A$, ${\bar q}^A$, see for example\cite{bib:eskola2,bib:eskola3}.
%
\section{Photon Structure}
\label{sec:phst}
 The investigation of the structure of the photon is a very active field 
 of research at LEP as well as at HERA, see\cite{bib:nisius} for a recent 
 review.
 At this conference new results from LEP and HERA have been
 presented which are briefly summarised here.
 \par
\begin{figure}[htb]
\begin{center}
{\includegraphics[width=0.70\linewidth]{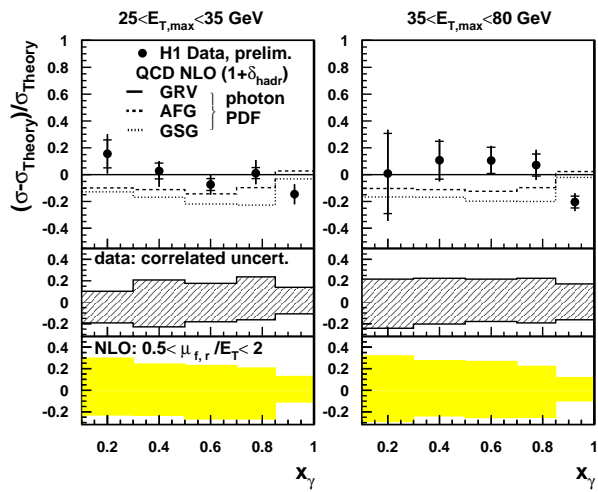}}
\caption{Measurement of the structure of quasi-real photons
         from H1.\label{fig:rn01}}
\end{center}
\end{figure}

 New investigations on the structure of the photon based on jet production 
 were presented by H1 for quasi-real\cite{bib:ferron}, and by ZEUS 
 for virtual\cite{bib:kcira} photons.
 The new H1 result, Figure~\ref{fig:rn01}, is consistent with the predictions 
 based on existing parametrisations of the hadronic photon structure function 
 $F_2^\gamma$ and at present the data are not precise enough to distinguish 
 between different parametrisations.
 This has to be confronted with the earlier result from ZEUS\cite{bib:ZEUS}
 which suggested that the parametrisations of $F_2^\gamma$, obtained from 
 fits to measurements made at e$^+$e$^-$ colliders, are too low for medium
 values of Bjorken $x$ and at factorisation scales of several hundred GeV$^2$.
%
%
 There are several differences between the ZEUS and H1 analysis such as the 
 choice made for the value of $\alpha_s$, the parton distributions used for 
 the photon, and most notably the difference in the corrections 
 applied to the data. 
 The H1 data are corrected for detector as well as hadronisation effects and
 are shown at the partonic level. In contrast, the ZEUS results are
 corrected only for detector effects and phase space regions are selected,
 where the hadronisation corrections, as implemented in Monte Carlo models,
 are found to be small.
 It remains to be seen how much of the apparent differences between the
 results can be explained by the different analysis methods.
 \par
\begin{figure}[htb]
\begin{center}
{\includegraphics[width=0.70\linewidth]{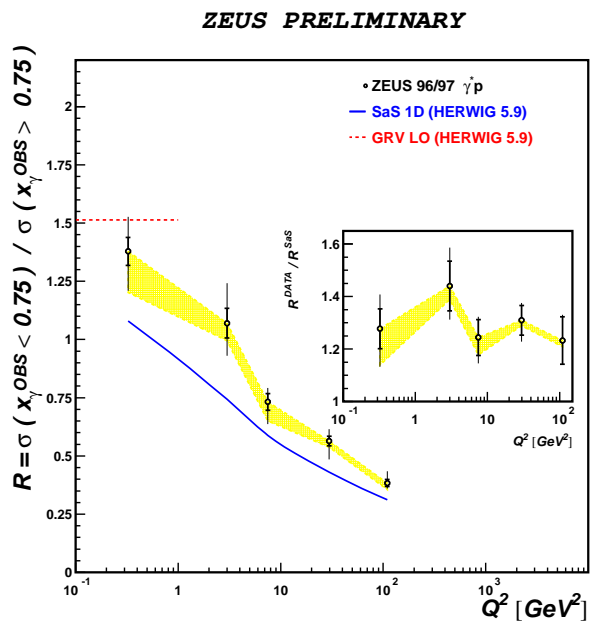}}
\caption{Measurement of the structure of virtual photons 
         from ZEUS.\label{fig:rn02}}
\end{center}
\end{figure}

 Also for the measurement of the structure of virtual photons some
 clarification is needed. 
 The recent ZEUS measurement, Figure~\ref{fig:rn02}, indicates some 
 shortcomings of the prediction from the SaS\cite{bib:SaS} parametrisations.
%
%
 The suppression of the photon structure with the photon virtuality
 is studied based on the ratio of the cross sections for low and large
 values of $x$. The SaS1D prediction fails to describe this ratio.
 Again, there seems to be a difference between the ZEUS result and the 
 earlier result from H1 on the structure of virtual photons\cite{bib:H1}.
 The H1 result, which is valid for an almost identical phase space region, 
 showed good agreement between the data and the predictions based on
 the GRV parametrisation of $F_2^\gamma$ with a virtuality suppression
 as given by the Drees-Godbole scheme.
 \par
 It is certainly desirable to complement the measurements of $F_2^\gamma$ 
 with the jet measurements from HERA which extend to larger 
 factorisation scales, when fits for the parton distribution
 functions of the photon are performed. 
 However, first it has to be seen if a consistent picture 
 of the various HERA results can be established.
%
\begin{figure}[htb]
\begin{center}
{\includegraphics[width=0.70\linewidth]{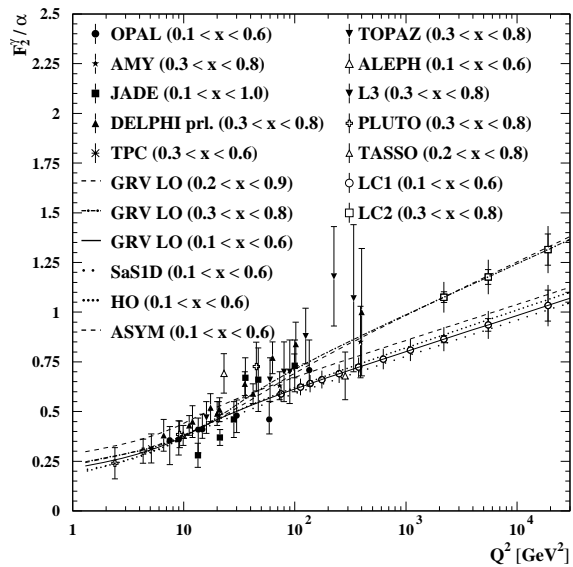}}
\caption{Prospects for the measurement of the $F_2^\gamma$ at a future linear 
         collider, from~\protect\cite{bib:nisius}.\label{fig:rn03}}
\end{center}
\end{figure}
%
 \par
 Recently at LEP progress has been made in the measurements of 
 $F_2^\gamma(x,Q^2)$.
 The phase space of the measurement has been extended by 
 OPAL\cite{bib:taylor}, both to lower values of $x$ and to larger values 
 of $Q^2$.
 In addition, significant progress in reducing the systematic uncertainty
 has been achieved by using improved Monte Carlo models to describe
 the hadronic final state\cite{bib:lepwg}, as well as by utilising
 a more sophisticated unfolding procedure in two dimensions\cite{bib:OPAL}.
 In view of these improvements a re-analysis of all LEP data using the 
 newly available analysis tools would significantly improve on the precision
 of $F_2^\gamma$.
 \par
 The prospects of future investigations of the photon structure 
 in the context of the planned linear collider programme are very 
 promising\cite{bib:deroeck}.
 The e$^+$e$^-$ linear collider will extend the available phase space,
 as shown e.g.~in Figure~\ref{fig:rn03} for the measurement of the $Q^2$ 
 evolution of $F_2^\gamma$ at medium $x$.
 The higher beam energy and luminosity compared to LEP also allows
 for the investigation of novel features like the measurement of
 the flavour dependence of $F_2^\gamma$ by exploring the exchange of $Z$ and
 $W$ bosons\cite{bib:zerwas}.
 When using highly polarised beams the measurement of structure 
 functions can be extended to polarised photons\cite{bib:stratmann}.
 The investigation of the photon structure will enter a completely new level
 of sophistication if a photon collider can be realised\cite{bib:deroeck}.
 Then photons of large energy with a rather moderate energy spread could
 be brought into collision, instead of the presently used soft photons from
 the Bremsstrahlungs spectrum at electron-positron colliders. 
%
%
\section{Proton structure}
\label{Sect:Prot}

\subsection{Deep inelastic scattering at high-$Q^2$}
\label{Sect:HiQ2DIS}

\begin{figure}[htb]
  \begin{center}
    \includegraphics[width=0.75\linewidth]{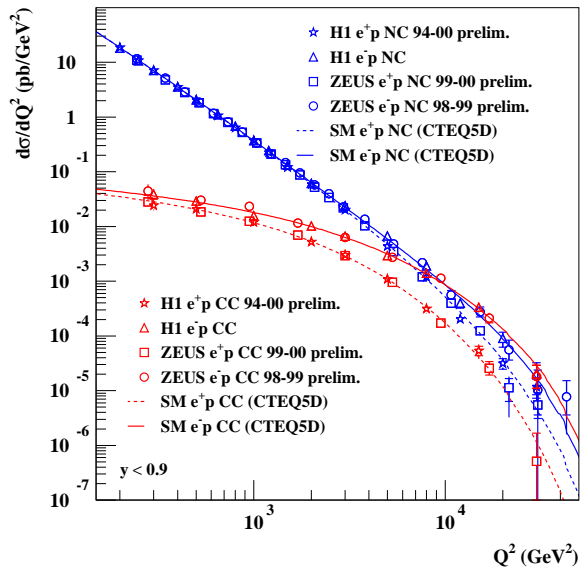}
    \caption{
      The high-$Q^2$ $e^\pm p$ neutral current and charged current
      deep inelastic scattering cross sections measured by the
      ZEUS and the H1 collaborations (points). 
      The predictions of the Standard Model evaluated with the CTEQ5D
      parton density functions are shown as the lines.
             }
    \label{Fig:NCCCdQ2}
  \end{center}
\end{figure}

New measurements of the high-$Q^2$ neutral current (NC) and charged
current (CC) deep inelastic scattering (DIS) $e^\pm p$ cross sections
were presented by the ZEUS\cite{Bib:Lopez} and H1\cite{Bib:Dubak}
collaborations \cite{Bib:ZH1HighQ2}.
Figure \ref{Fig:NCCCdQ2} shows the differential cross section 
$d\sigma / dQ^2$ for NC DIS.
The ZEUS  $e^+p$ data, presented for the first time at this conference,
were obtained using approximately 60~pb$^{-1}$ of $e^+p$ data
collected at a centre-of-mass energy of $\sqrt{s}=318$~GeV over the
years 1999 and 2000. 
H1 has combined 46~pb$^{-1}$ of $e^+p$ data 
(from a total of 65~pb$^{-1}$ collected in 1999 and 2000) at
$\sqrt{s}=318$~GeV with the 38~pb$^{-1}$ of data collected over the
years 1994 to 1997 at $\sqrt{s}=300$~GeV.
The NC cross section is observed to fall by roughly 6 orders of
magnitude as $Q^2$ goes from $Q^2 \sim 200$~GeV$^2$ to 
$Q^2 \sim 10\,000$~GeV$^2$. 
For $Q^2 < 1\,000$~GeV$^2$ the NC $e^+p$ cross section is observed to
be approximately equal to the $e^-p$ cross section.
In this kinematic domain NC $e^\pm p$ DIS is mediated by single photon
exchange and the $Q^2$ dependence of $d\sigma / dQ^2$ reflects the
$1/Q^4$ dependence of the photon propagator.
As $Q^2$ approaches the $Z$-boson mass squared ($Q^2 \sim M_Z^2$) the
$e^+p$ cross section falls below that for $e^-p$ NC DIS.
At such large $Q^2$ both photon- and $Z$-exchange contributions must
be taken into account.
The interference between the two contributions suppresses the $e^+p$
cross section and enhances the $e^-p$ cross section.
The Standard Model (SM), which incorporates all these effects, evaluated 
with the CTEQ5\cite{Bib:CTEQ5} parton density functions (PDFs) gives a 
good description of the data.

The differential cross section $d\sigma / dQ^2$ for CC $e^\pm p$ DIS
is shown in Figure \ref{Fig:NCCCdQ2}. 
The cross sections fall slowly for $200<Q^2<1\,000$~GeV$^2$ and then
more rapidly as $Q^2$ approaches the $W$-boson mass squared
($Q^2 \sim M_W^2$).
The SM, evaluated with the CTEQ5 PDFs, gives a good description of the
data.
The $e^-p$ CC DIS cross section is always larger than the $e^+p$ one. 
The CC interaction distinguishes between quarks of different flavour. 
Thus, the $e^-p$ CC process picks out the up-type quarks and the
down-type anti-quarks while $e^+p$ CC DIS picks out the down-type
quarks and the up-type anti-quarks.
The $e^-p$ cross section is larger than the $e^+p$ cross section
because there are more $u$-valence quarks than $d$-valence quarks in
the proton and because the chiral nature of the CC interaction
suppresses the contribution of the down-type quarks in $e^+p$ CC DIS. 
Note that for $Q^2>10\,000$~GeV$^2$ the four cross sections for NC and
CC DIS in $e^+p$ and $e^-p$ collisions all have a similar magnitude.
This is a striking experimental verification of the electroweak
unification embodied in the Standard Model.

\begin{figure}[htb]
  \begin{center}
    \includegraphics[width=0.9\linewidth]{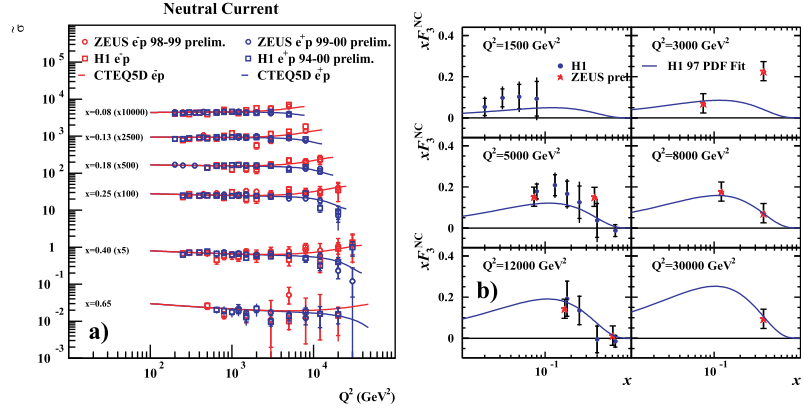}
    \caption{
      (a) The reduced cross section for $e^\pm p$ neutral current deep 
          inelastic scattering measured by the ZEUS and the
          H1 collaborations (points).
          The predictions of the Standard Model evaluated with the
          CTEQ5D parton density functions are shown as the lines. 
      (b) The neutral current structure function $xF_3^{\rm NC}$
          extracted from $e^\pm p$ neutral current deep inelastic
          scattering data by the ZEUS and H1 collaborations (points). 
          The predictions of the Standard Model evaluated with the
          CTEQ5D parton density functions are shown as the solid
          lines.
             }
    \label{Fig:NCRedXSect}
  \end{center}
\end{figure}

The reduced cross section $\tilde{\sigma}^{e^\pm p}_{\rm NC}$ is
related to the double differential $e^\pm p$ NC DIS cross section by
\begin{equation}
  Y_+ \tilde{\sigma}^{e^\pm p}_{\rm NC} =
  \left[ \frac{2 \pi \alpha^2}{Q^4 x} \right]^{-1}
  \frac{d^2 \sigma ^{e^\pm p}_{\rm NC}}{dx dQ^2} =
  Y_+ F_2^{\rm NC} \mp Y_- xF^{\rm NC}_3 - y^2 F^{\rm NC}_L
  \label{Eq:NCBorn}
\end{equation}
where $Y_\pm = 1 \pm (1-y)^2$ and $y$ is related to $Q^2$ and $x$ by
$Q^2 = xys$.
Figure \ref{Fig:NCRedXSect}a shows $\tilde{\sigma}^{e^\pm p}_{\rm NC}$
plotted as a function of $Q^2$ for several values of $x$.
For $Q^2 < 1\,000$~GeV$^2$ and $x \sim 0.1$ the reduced cross section
is observed to be almost independent of $Q^2$ while for 
$Q^2 < 1\,000$~GeV$^2$ and $x > 0.25$ it falls slowly with $Q^2$.
For $Q^2$ above $1\,000$~GeV$^2$ the $e^-p$ reduced cross section
exceeds the $e^+p$ reduced cross section. 
The Standard Model (SM) expectations for $\tilde{\sigma}^{e^\pm p}_{\rm NC}$ 
obtained by evaluating Equation \ref{Eq:NCBorn} using standard parton
density functions gives a good description of the data.
The SM accounts for the difference between $\tilde{\sigma}^{e^-p}_{\rm NC}$ 
and $\tilde{\sigma}^{e^+ p}_{\rm NC}$ in terms of the structure
function $xF^{\rm NC}_3$ so that $xF^{\rm NC}_3$ may be extracted from
the difference between $\tilde{\sigma}^{e^-p}_{\rm NC}$ 
and $\tilde{\sigma}^{e^+ p}_{\rm NC}$.
Both ZEUS and H1 have extracted $xF^{\rm NC}_3$ in this way, the
results are plotted in Figure \ref{Fig:NCRedXSect}b.
A good description of the data is obtained when $xF_3^{\rm NC}$ is
evaluated with standard PDFs.
More $e^-p$ data is required to improve the precision of 
$xF^{\rm NC}_3$.
With sufficient data $xF^{\rm NC}_3$ will be used to constrain the
valence quark PDFs.

\begin{figure}[htb]
  \begin{center}
    \includegraphics[width=0.475\linewidth]{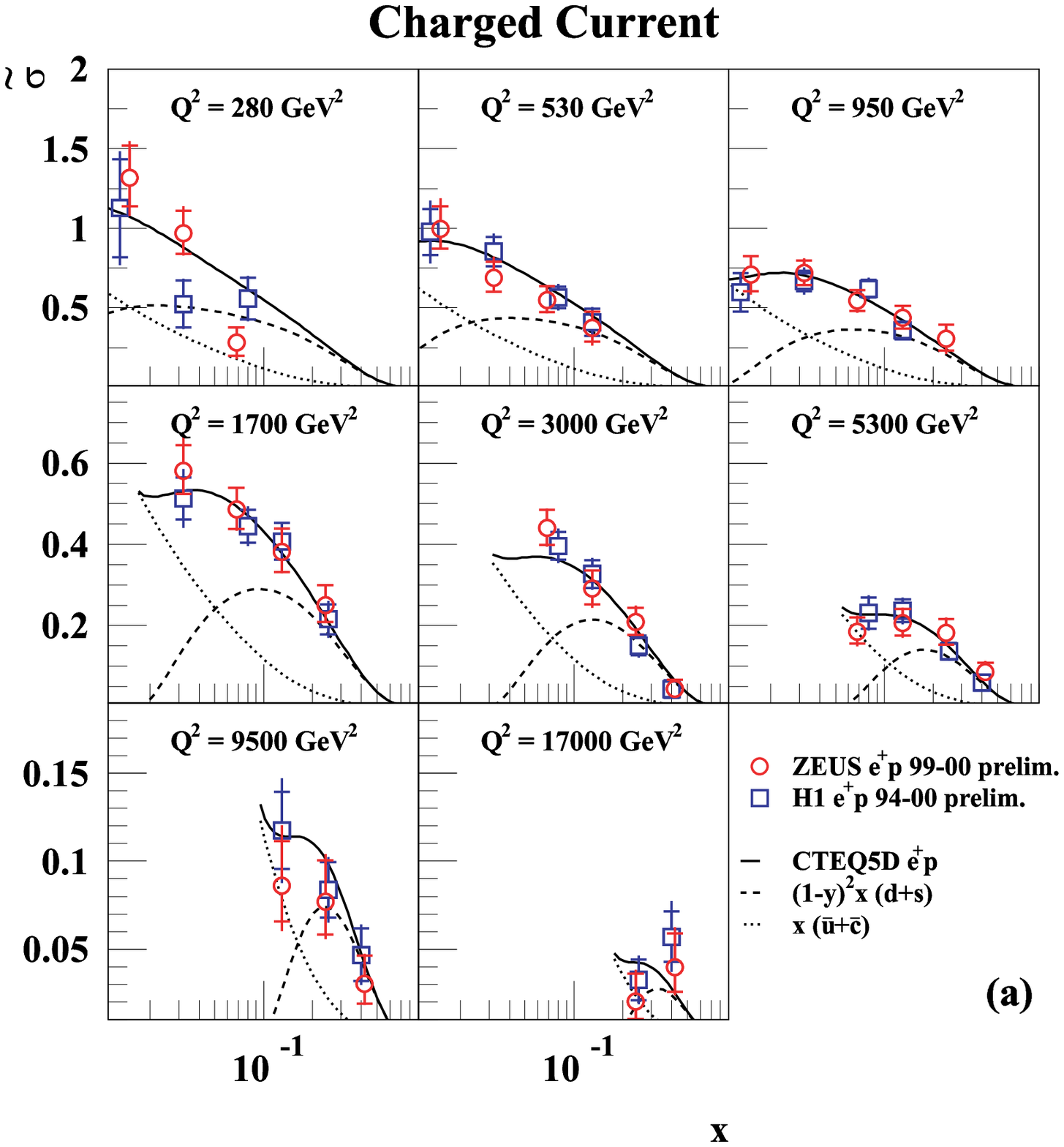}
    \includegraphics[width=0.475\linewidth]{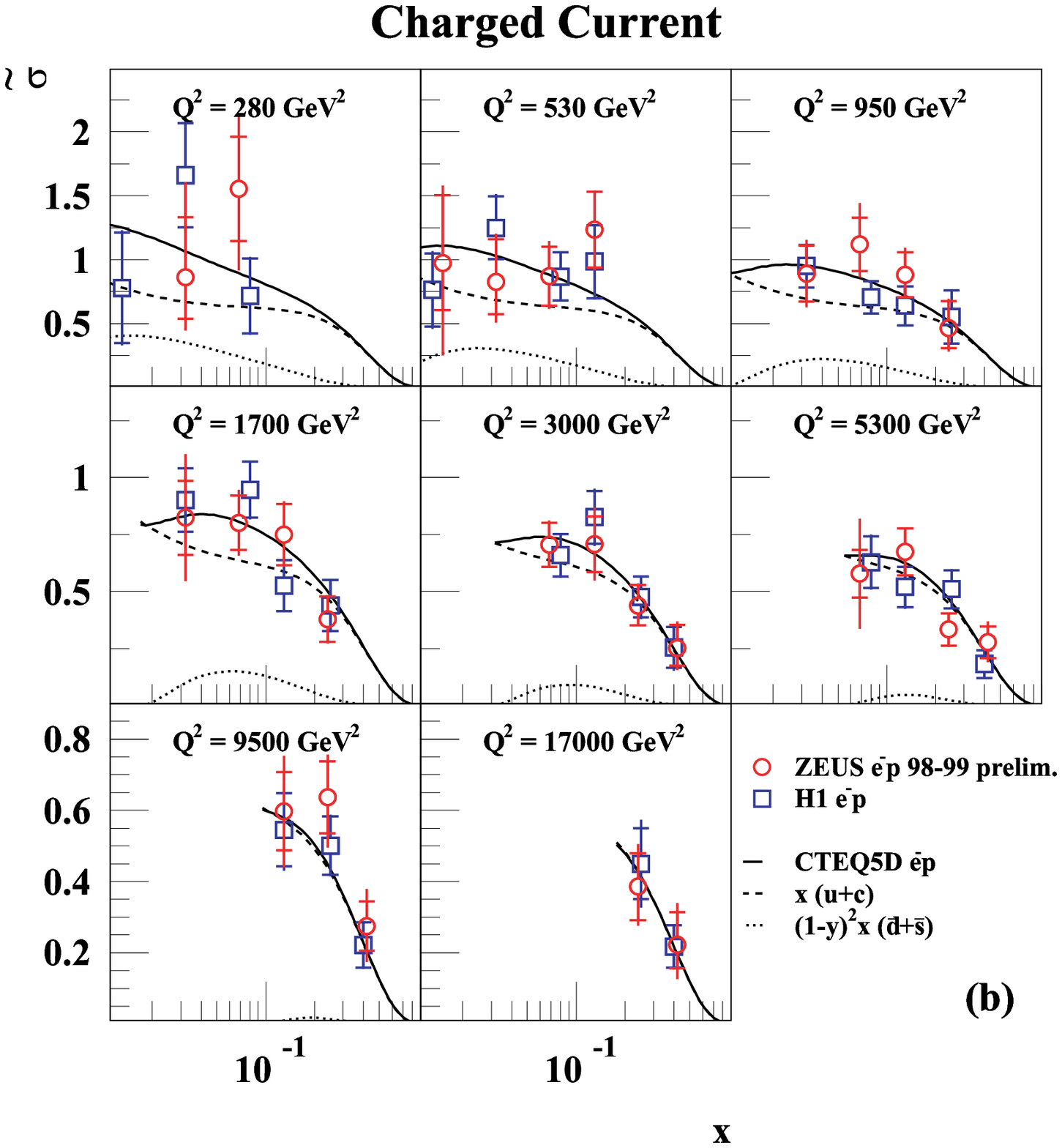}
    \caption{
      The reduced cross section for $e^+ p$ (a) and $e^- p$ (b)
      charged current deep inelastic scattering measured by the ZEUS
      and the H1 collaborations (points).
      The predictions of the Standard Model evaluated with the
      CTEQ5D parton density functions are shown as the solid lines. 
      Quark contributions, evaluated using the CTEQ5D PDFs, are shown
      as the dashed lines.
             }
    \label{Fig:CCRedXSect}
  \end{center}
\end{figure}

The charged current can also be used to study the partonic content of
the proton. 
The reduced cross section for $e^\pm p$ CC DIS, 
$\tilde{\sigma}^{e^\pm p}_{\rm CC}$, is defined to be
\begin{equation}
  \tilde{\sigma}^{e^\pm p}_{\rm CC} = 
  \left[ 
    \frac{G^2_F}{2 \pi x} \frac{M_W^4}{\left( M_W^2 + Q^2 \right)^2 }
  \right]^{-1}
  \frac{d^2 \sigma ^{e^\pm p}_{\rm CC}}{dx dQ^2} =
  \frac{1}{2} 
  \left[
    Y_+ F_2^{\rm CC} \mp Y_- xF^{\rm CC}_3 - y^2 F^{\rm CC}_L
  \right]
  \label{Eq:CCRedXSect}
\end{equation}
Figure \ref{Fig:CCRedXSect} shows $\tilde{\sigma}^{e^\pm p}_{\rm CC}$
as a function of $x$ for values of $Q^2$ between 200~GeV$^2$ and
$20\,000$~GeV$^2$.
The SM expectation of Equation \ref{Eq:CCRedXSect} gives a good
description of the data.
At leading order in QCD $\tilde{\sigma}^{e^+ p}_{\rm CC}$ is given by
\begin{equation}
  \tilde{\sigma}^{e^+ p}_{\rm CC} = x\left[ 
                      \bar{u} + \bar{c} + (1 - y)^2 (d + s)
                                     \right]
  \label{Eq:CCQPMPos}
\end{equation}
while $\tilde{\sigma}^{e^- p}_{\rm CC}$ is given by
\begin{equation}
  \tilde{\sigma}^{e^+ p}_{\rm CC} = x\left[
                   u + c + (1 - y)^2 (\bar{d} + \bar{s})
                                     \right]
  \label{Eq:CCQPMEle}
\end{equation}
Figure \ref{Fig:CCRedXSect} shows how the different quark flavours
contribute to the reduced cross section. 
It can be seen that, at high-$x$, $\tilde{\sigma}^{e^+ p}_{\rm CC}$
is sensitive to the $d$-quark density while 
$\tilde{\sigma}^{e^- p}_{\rm CC}$ is sensitive to the $u$-quark
density. 

These results, important in themselves, represent a significant
milestone in the development of the HERA physics programme.
DIS01 took place at a time when major upgrades to the HERA machine
and experiments were nearing completion\cite{Bib:Elsen,Bib:Foster}.
The HERA machine upgrade will provide polarised electron and positron
beams for the colliding beam experiments ZEUS and H1 at a luminosity
five times greater than has been achieved to date.
The new high-$Q^2$ cross section results, therefore, represent the
culmination of the programme of measurements of the unpolarised DIS
cross sections at high-$Q^2$.

\subsection{Determination of $\alpha_{\rm S}$ and extraction of the
            gluon density} 
\label{Sect:AlpSandxG}

A highlight of the structure function session was the variety of new,
high precision, simultaneous determinations of the strong coupling
constant, $\alpha_{\rm S}$, and the gluon density, $xG$.
The ZEUS\cite{Bib:Nagano} and H1\cite{Bib:Wallny} collaborations
presented new analyses in which the scaling violations of the structure
function $F_2^{\rm NC}$ are exploited to determine $xG$ and
$\alpha_{\rm S}$.
Both collaborations have performed a NLO QCD fit using their own data on
$\tilde{\sigma}^{e^+ p}_{\rm NC}$ together with structure function
data from other experiments.
The analyses proceed by parameterising the PDFs at a starting scale
$Q^2_0$.
The DGLAP equations\cite{Bib:DGLAP} are used to evolve
from $Q^2_0$ to the value of $Q^2$ that corresponds to a particular
data point. 
The relevant measured quantity is then determined from the evolved PDFs
and the contribution to the total $\chi^2$ calculated.
The ZEUS and H1 analyses differ both in philosophy and in the details
of the fitting technique used.

The goal of the H1 collaboration was to determine $xG$ and
$\alpha_{\rm S}$ as precisely as possible using as few data sets as
possible.
The fit includes data from H1\cite{Bib:H1F2} and
BCDMS\cite{Bib:BCDMSF2} (proton target data) in order that the lever
arm in $x$ and $Q^2$ be sufficient to constrain the parameterisation. 
The PDFs are parameterised at $Q^2_0 = 4$~GeV$^2$.
The PDF parameterisation is based on effective valence and sea
distributions together with a parameterisation for $xG$.
The number of parameters in the gluon and sea distributions are chosen
to saturate the $\chi^2$.
The $c$- and $b$-quark contributions to the cross sections are
evaluated using photon-gluon fusion using a massive fixed flavour
number scheme\cite{Bib:FFN}.

ZEUS parameterised the $u$- and $d$-valence,  the sea and the gluon
PDFs at $Q^2_0 = 7$~GeV$^2$. 
In addition, the difference between the $u$- and $d$-quark density
($x(u-d)$) was parameterised.
The recent ZEUS data\cite{Bib:ZEUSF2} was used together with data from 
BCDMS\cite{Bib:BCDMSF2}, CCFR\cite{Bib:CCFRxF3}, NMC\cite{Bib:NMC} and
E665\cite{Bib:E665}. 
A modified massive variable flavour number scheme (RT-VFN
scheme\cite{Bib:RTVFN}) was used.
This treatment of the charm quark smoothly interpolates between the
low-$Q^2$ region where threshold effects are important to the high-$Q^2$
region where the charm quark mass is negligible.

\begin{figure}[htb]
  \begin{center}
    \includegraphics[width=0.9\linewidth]{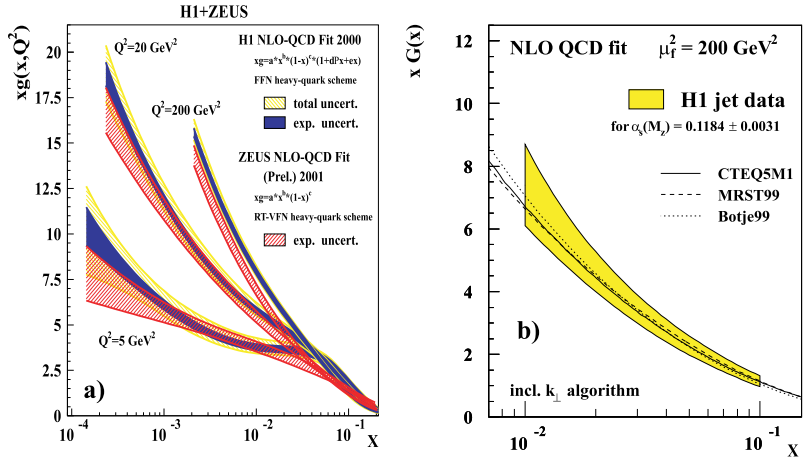}
    \caption{
      (a) The gluon distributions extracted from the ZEUS and H1 NLO
      QCD fits to structure function data.
      The statistical and systematic error bands are shown shaded.
      Details of the extraction of these error bands may be found
      elsewhere in these proceedings.
      (b) The gluon distribution extracted from a NLO QCD fit to the 
      H1 inclusive jet cross section data.
             }
    \label{Fig:ZH1Gluon}
  \end{center}
\end{figure}

Both ZEUS and H1 have incorporated a very careful treatment of the
experimental systematic errors which includes the point-to-point
correlations.
H1 allowed the fit to determine the best values for the
various systematic uncertainties, the ZEUS fit did not.
The resulting gluon distributions are compared in Figure
\ref{Fig:ZH1Gluon}a.
When the differences between the two analyses are taken into account
the agreement between the two results is reasonable.
The corresponding values of $\alpha_{\rm S}$ are
\begin{eqnarray}
  \alpha_{\rm S} & = & 0.1172 \pm 0.0008 {\rm (stat. + uncor.~sys.)}
                              \pm 0.0054 {\rm (cor. sys.)}
                              {\rm \,\,\,\,(ZEUS)}              \\
  \alpha_{\rm S} & = & 0.1150 \pm 0.0017 {\rm (expt.)}
                              ^{+0.0009}_{-0.0005} {\rm (model)}
                              \pm 0.005 {\rm (scale)}
                              {\rm \,\,\,\,(H1)}
\end{eqnarray}
ZEUS breaks up the total experimental uncertainty into the contribution
arising from the correlated experimental systematic uncertainties
(cor. sys.) and the combined statistical and uncorrelated
systematic uncertainties (${\rm stat. + uncor.~sys.}$).
H1 has combined the statistical and experimental systematic
contributions into a single error labeled (expt.). 
The H1 `model' error accounts for uncertainties in the theoretical
model such as those arising from the choice of the functional form or 
the treatment of heavy quarks while the `scale'
error is evaluated by varying the factorisation and renormalisation
scales. 
The values obtained are consistent with one another and with the values
obtained in various global analyses\cite{Bib:GlobAlphaS}.
Given the importance of the measurements it will be important to
understand in detail the range of applicability of the results and, if
possible, to converge upon a common approach.
\begin{figure}[htb]
  \begin{center}
    \includegraphics[width=0.475\linewidth]{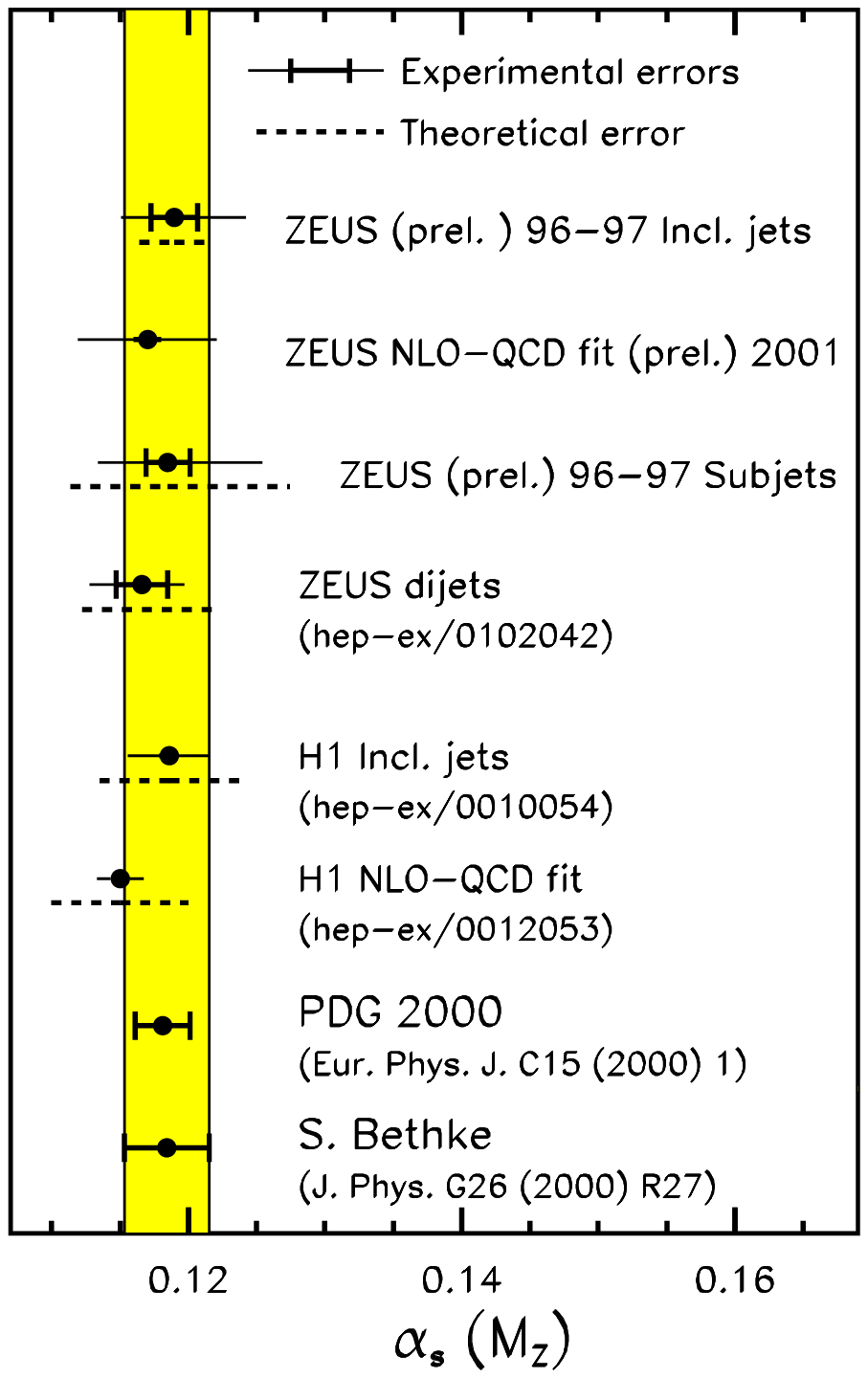}
    \caption{
      Compilation of the various determinations of the strong coupling
      constant, $\alpha_S$, at HERA together the results of two global
      analyses.
             }
    \label{Fig:ZH1AlphaSSummary}
  \end{center}
\end{figure}
The ZEUS\cite{Bib:Tassi} and H1\cite{Bib:Grindhammer} collaborations
have exploited the sensitivity of DIS jet production cross sections to
determine $\alpha_{\rm S}$ and $xG$.
With $\alpha_{\rm S}$ set equal to the world average, $xG$ may be
determined in a fit to the measured inclusive jet cross sections.
This has been done by H1 and the result is shown in Figure
\ref{Fig:ZH1Gluon}b\cite{Bib:H1JetGlu}.
Alternatively, the PDFs can be taken from one of the standard
parameterisations and the jet cross sections used to determine
$\alpha_{\rm S}$. 
When $\alpha_{\rm S}$ is determined in this way it is essential to
estimate the uncertainty in $\alpha_{\rm S}$ which arises from the
choice of PDF. 
Both ZEUS\cite{Bib:ZEUSJetAlphaS} and H1\cite{Bib:H1JetGlu} have
performed such analyses.
The H1 analysis uses the inclusive jet cross section.
ZEUS has chosen to analyse the ratio of the cross section for the
production of 2-jets (plus the target remnant) to the total inclusive
DIS cross section.
The use of the ratio reduces the sensitivity of the result to the
choice of PDF.
A compilation of the values of $\alpha_{\rm S}$ obtained from jet 
production data is shown in Figure \ref{Fig:ZH1AlphaSSummary} together 
with the results obtained from the scaling violations of $F_2$.
The results agree with one another and with the world average.
The way in which $\alpha_{\rm S}$ depends on $Q^2$, the running of
$\alpha_{\rm S}$, may also be investigated using the jet data.
Both ZEUS\cite{Bib:ZEUSJetAlphaS} and H1\cite{Bib:H1JetGlu} have
performed such analyses and the results are consistent with the QCD
expectation. 
Finally, it is also possible to estimate $xG$ and $\alpha_{\rm S}$
simultaneously from the jet data in a fit conceptually similar to that
used in the analysis of the scaling violations of $F_2^{\rm NC}$
described above. 
The H1\cite{Bib:H1JetGlu} collaboration has performed such an
analysis. 
The fit results in estimates of the gluon distribution and
$\alpha_{\rm S}$ which are correlated.

\begin{figure}[htb]
  \begin{center}
    \includegraphics[width=0.475\linewidth]{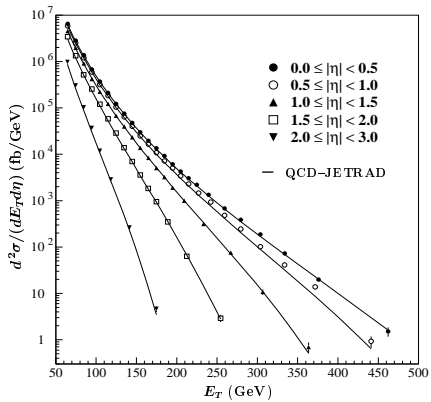}
    \caption{
      Inclusive double differential single inclusive jet cross
      sections measured by the D0 collaboration (points) for
      jet pseudo-rapidity ($\eta$) intervals plotted as a
      function of the jet transverse energy ($E_T$).
      The Standard Model prediction is shown as the solid line.
             }
    \label{Fig:D0Jets}
  \end{center}
\end{figure}

The techniques discussed above do not constrain the gluon distribution
for $x>0.1$. 
A promising new development is the measurement of the double
differential inclusive jet cross section $d^2 \sigma / dE_T d\eta$
(where $E_T$ is the jet transverse energy and $\eta$ is the jet
pseudo-rapidity) by the D0 collaboration\cite{Bib:Babukhadia,Bib:D0Jets}.
The cross sections, shown in Figure \ref{Fig:D0Jets}, have been
measured over the kinematic range $| \eta | < 3$ and
$50<E_T<500$~GeV.
Since the beam energy at the Tevatron is 900~GeV, these cross sections
are sensitive to $xG$ for values of $x$ as high as $x\sim 0.9$.
The sensitivity of these cross sections to the gluon was demonstrated
by evaluating a $\chi^2$ formed between the measurements and the
inclusive jet cross section evaluated using various standard PDF
sets. 
The $\chi^2$ evaluation took account of correlations between
uncertainties in the variables $\eta$ and $E_T$.
The results indicate that the data prefer a slightly harder
gluon distribution than is contained within the standard 
CTEQ4\cite{Bib:CTEQ4} parameterisation. 
NLO QCD fits incorporating the D0 inclusive jet data are being
prepared by the MRST and the CTEQ groups.
The data will provide a stringent constraint on the gluon density at
high-$x$.

\subsection{Structure functions at low-$Q^2$}
\label{Sect:F2LowQ2}

\begin{figure}[htb]
  \begin{center}
    \includegraphics[width=0.475\linewidth]{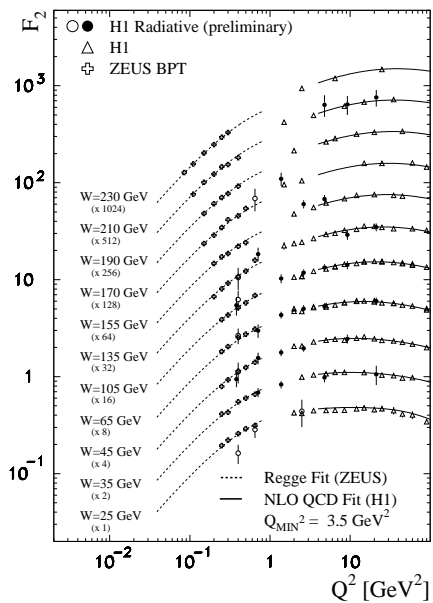}
    \caption{
      Compilation of ZEUS and H1 measurements of $F_2$ at
      low-$Q^2$ in NC DIS (points) plotted at several values of 
      $W$ as a function of $Q^2$.
      The solid lines show the result of the H1 NLO QCD fit to the
      H1 $F_2$ data for $Q^2>3.5$~GeV$^2$.
      The dashed lines show the result of the ZEUS Regge fit to the
      ZEUS $F_2$ data for $Q^2<0.65$~GeV$^2$.
             }
    \label{Fig:ZH1LowQ2F2}
  \end{center}
\end{figure}

It is well established that $F_2^{\rm NC}$ rises rapidly as $x$ falls
below $x \sim 0.01$ so long as $Q^2$ is larger than $\sim 2$~GeV$^2$
and that NLO QCD gives a good description of the data for
$Q^2>2$~GeV$^2$~\cite{Bib:H1F2,Bib:ZEUSF2}. 
Precise measurements of $F_2^{\rm NC}$ for $Q^2$ in the range
$0.045<Q^2<0.65$~GeV$^2$ show that in this kinematic domain $F_2^{\rm NC}$
does not rise rapidly as $x$ falls\cite{Bib:ZEUSBPTF2}.
The low $Q^2$ data can be described using non-perturbative models.
The $x$ dependence of $F_2^{\rm NC}$ appears to undergo a transition
at a $Q^2$ of around 1~GeV$^2$.
The nature of the transition between the kinematic domain where
perturbative QCD is applicable and the domain where non-perturbative
strong interaction physics is important was discussed in a dedicated
session at this workshop.
The contributions to this session are summarised elsewhere in these
proceedings\cite{Bib:Bartels}.
A new measurement of $F_2^{\rm NC}$ spanning the $Q^2$ range
$0.35<Q^2<20$~GeV$^2$ was presented by the H1
collaboration\cite{Bib:Issever}.
The low values of $Q^2$ were accessed by selecting a sample of events
in which a photon was radiated in the direction of the incoming
positron. 
Such initial state radiation (ISR) causes the effective centre-of-mass
energy in the $e^+p$ collision to be reduced.
Hence, a scattered positron of particular energy and scattered
through a particular angle results from a DIS collision is characterised
by a smaller value of $Q^2$ and a higher value of $x$ than would be
the case if a scattered positron with the same energy and scattering
angle were found in the absence of ISR.
Figure \ref{Fig:ZH1LowQ2F2} shows the resulting $F_2$ plotted as a
function of $Q^2$ at several values of $W$, the centre-of-mass energy in
the virtual photon-proton system.
The new H1 ISR measurement spans the transition region around 
$Q^2 \sim 1$~GeV$^2$.
Further measurements in this important kinematic region will be
required in order that an understanding of the
underlying physics may be developed.

The CCFR collaboration has resolved the long-standing discrepancy
between the structure functions extracted from neutrino-nucleon CC DIS
and those extracted from muon-nucleon NC
DIS\cite{Bib:Bernstein,Bib:CCFRReAnal}. 
The expression for the cross sections for CC $\nu N$ DIS contains three
structure functions, just as Equation \ref{Eq:CCRedXSect} does, but
the quark composition of these structure functions in the Quark Parton
Model differs from those given in Equations \ref{Eq:CCQPMPos} 
and \ref{Eq:CCQPMEle}\cite{Bib:Devenish}.
The CCFR data were re-analysed\cite{Bib:Bernstein,Bib:CCFRReAnal}
taking charm mass effects into account when evaluating NLO QCD
corrections for the longitudinal structure function and the difference
between the structure function $xF_3$ for $\nu N$ and $\bar{\nu} N$
CC DIS. 
In addition to an update of the structure function $F_2^\nu$ for
$Q^2>1$~GeV$^2$, the CCFR collaboration presented results on $F_2^\nu$
for $Q^2<1$~GeV$^2$ \cite{Bib:Bernstein,Bib:CCFRF2}. 
This is the first time that $F_2^\nu$ has been determined for such low
values of $Q^2$.

\subsection{The longitudinal structure function, $F_{\rm L}^{\rm NC}$}
\label{Sect:ZH1FL}

\begin{figure}[htb]
  \begin{center}
    \includegraphics[width=0.475\linewidth]{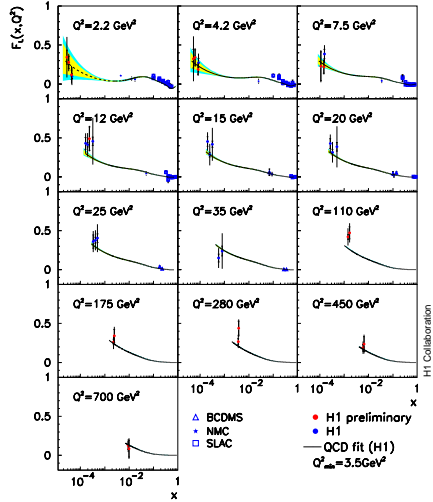}
    \caption{
      The longitudinal structure function, $F_L$, extracted from 
      NC $ep$ DIS data by the H1 collaboration.
             }
    \label{Fig:ZH1FL}
  \end{center}
\end{figure}

The structure function $F_{\rm L}^{\rm NC}$ is multiplied by a factor of
$y^2$ in the expression for the cross section for NC $e^+ p$ DIS.
This makes $F_{\rm L}^{\rm NC}$ difficult to determine at HERA since
the NC cross section $d^2 \sigma^{e^+p}_{\rm NC} / dx dy \propto
1/y^2$ at fixed $x$.
A direct determination of $F_{\rm L}^{\rm NC}$ can be made by reducing
the proton beam energy and so raising the value of $y$ for fixed $x$
and $Q^2$~\cite{Bib:HERAWSFL}. 
The difference between the proton beam energies at which HERA has
operated to date (820~GeV and 920~GeV) is too small to allow an
extraction of $F_{\rm L}^{\rm NC}$.
H1 has developed two methods of extracting $F_{\rm L}^{\rm NC}$
\cite{Bib:H1F2}.
In the first, indirect, method a NLO QCD fit was made to the reduced
cross section for $y<0.3$ where the contribution of $F_{\rm L}^{\rm NC}$ 
may be neglected.
This fit was then extrapolated to larger values of $y$ and the
difference between the extrapolation and the data taken as a measure
of $F_{\rm L}^{\rm NC}$.
For $Q^2$ less than 10~GeV$^2$ $d\sigma^{e^+p}_{\rm NC} /d\ln y$ was
used instead of the NLO QCD fit to extrapolate to 
high-$y$\cite{Bib:Eckstein}. 
Secondly, the ISR technique, outlined above, has been used to vary
the effective centre-of-mass energy so that $F_{\rm L}^{\rm NC}$ can
be extracted.
Both methods rely upon the measurement of the reduced cross section
at the highest possible $y$. 
At DIS01 the H1 collaboration presented new data on
$\tilde{\sigma}^{e^+ p}$ for $1.5<Q^2<12$~GeV$^2$ from a sample that
contains scattered positrons with energies as low as 3~GeV.
As a result $y$ values as high as 0.9 can be accessed.
In addition, H1 has extracted $F_{\rm L}^{\rm NC}$ for $Q^2$ values as
high as 700~GeV$^2$~\cite{Bib:Dubak}.
Figure \ref{Fig:ZH1FL} shows the values of $F_{\rm L}^{\rm NC}$
extracted using these methods as well as the $F_{\rm L}^{\rm NC}$
resulting from the fit to the $F_2$ data for $y<0.3$.

The structure function $R^\nu$ 
($F^\nu_{\rm L} \approx \left[ R^\nu / (1 + R^\nu) \right] F_2^\nu$)
extracted from $\nu N$ and $\bar{\nu} N$ data by the CCFR
collaboration has been measured\cite{Bib:Bodek}.
$R^\nu$ was presented over the kinematic range $0.004<x<0.5$ and
$0.5<Q^2<200$~GeV$^2$. 
The data are well described by NLO QCD evaluated using the MRST
PDFs\cite{Bib:MRST}. 

\subsection{Flavour specific parton density functions}
\label{Sect:FlavPDFs}

Much can be inferred about the partonic structure of the proton from
the inclusive measurements discussed above.
A deeper understanding can be developed using measurements of the
cross sections for processes that are sensitive to specific quark
flavours. 
The sensitivity of CC data from HERA to the $d$- and $u$-quark
densities was discussed in Section \ref{Sect:HiQ2DIS}.
An estimate of the PDF ratio $\bar{d}/\bar{u}$ can be obtained by
determining the cross section for the Drell-Yan processes
$p N \rightarrow \mu^+ \mu^- X$ where $N$ is either a proton or a
deuteron.
Final measurements of the proton-proton to proton-deuteron
Drell-Yan cross section ratio from the E866 experiment at Fermilab
were presented\cite{Bib:Isenhower}.
The Bjorken-$x$ of the two partons that annihilate, $x_1$ and $x_2$,
can be reconstructed from the combined four momentum of the muon
pair. 
When $x_1 >> x_2$ the cross section ratio is approximately given by
\begin{equation}
             \frac{
                    \left.
                          \frac{d^2\sigma^{pd}}{dx_1 dx_2} 
                    \right|_{x_1>>x_2}
                    }
                   {
                    \left.
                          \frac{d^2\sigma^{pp}}{dx_1 dx_2} 
                    \right|_{x_1>>x_2}
                    }
              \approx
                    \left[
                          1 + \frac{\bar{d}({x})}{\bar{u}({x})}
                    \right]
  \label{Eq:DrellYanRat}
\end{equation}
where $x = x_2$.
The $\bar{d}/\bar{u}$ ratio extracted from Equation
\ref{Eq:DrellYanRat} by the E866 collaboration is observed to
peak at $x \sim 0.18$\cite{Bib:Isenhower}. 
This measurement will give an important constraint on the sea quark
density at intermediate to high-$x$.

\begin{figure}[htb]
  \begin{center}
    \includegraphics[width=0.9\linewidth]{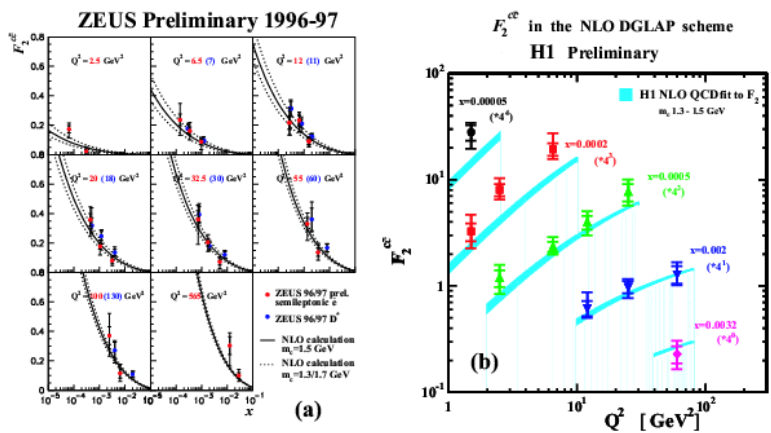}
    \caption{
      The charm contribution to $F_2^{\rm NC}$, $F_2^{c\bar{c}}$. 
      (a) $F_2^{c\bar{c}}$ plotted as a function of $x$ for several
      values of $Q^2$ (points).
      The solid line shows the result of a NLO QCD calculation.
      (b) $F_2^{c\bar{c}}$ plotted as a function of $Q^2$ for several
      values of $x$. 
      The expectations of the H1 NLO QCD fit are shown as the shaded
      bands.
             }
    \label{Fig:ZH1F2CC}
  \end{center}
\end{figure}

Charm production in DIS at HERA has been exploited by both the 
ZEUS\cite{Bib:Schagen} and the H1\cite{Bib:Mohrdieck} collaborations
to extract $F_2^{c\bar{c}}$, the charm contribution to $F_2^{\rm NC}$. 
To date two methods have been used to tag charm in DIS.
The first method, used by ZEUS and H1, exploits the small mass difference 
between the charmed mesons in the decay $D^* \rightarrow D^0 \pi$ to 
identify $D^*$ production.
The ZEUS collaboration has also used the electrons produced in
semi-leptonic charm decays.
The selection of the semi-leptonic sample was challenging and was
performed by combining information on the electromagnetic shower
profile in the calorimeter with the energy loss information obtained
from the central drift chamber.
The steps by which $F_2^{c\bar{c}}$ was extracted are common to both
collaborations.
First the DIS charm cross section times branching ratio was measured
in a region of phase space limited by the acceptance of the
apparatus.
Next, the cross section was extrapolated to the full phase space using
QCD based Monte Carlo programs and the known branching ratios of the
charmed hadrons taken into account.
Finally, $F_2^{c\bar{c}}$ was extracted from the charm production
cross section.
The results are plotted in Figure \ref{Fig:ZH1F2CC}.
$F_2^{c\bar{c}}$ is observed to rise rapidly as $x$ falls, a behaviour
that mirrors the behaviour of the inclusive $F_2^{\rm NC}$.
In addition, $F_2^{c\bar{c}}$ is observed to exhibit the strong scaling
violations expected in NLO QCD calculations of $F_2^{c\bar{c}}$ based
on standard PDFs.
$F_2^{c\bar{c}}$ is large, amounting to $\sim 25\%$ of $F_2^{\rm NC}$ for 
$Q^2 \sim 30$~GeV$^2$ and $x \sim 10^{-3}$.
The high luminosity that will be provided by the HERA upgrade 
coupled with the increased charm sensitivity offered by the new ZEUS
and H1 silicon micro-vertex detectors promises rapid progress in this
field. 

The CCFR and the NuTeV collaborations each presented several
new results that are sensitive to the flavour content of the proton.
NuTeV\cite{Bib:Goncharov} presented final cross sections for dimuon
production in $\nu N$ and $\bar{\nu} N$ DIS.
Dimuon production in CC $\nu N$ DIS is sensitive to the strange quark
density. 
By making a NLO QCD fit to the $\nu N$ and the $\bar{\nu} N$ dimuon
data NuTeV has been able to demonstrate that the magnitude of the
strange sea amounts to approximately 40\% of the up-quark sea and
that the $s$-quark density has approximately the same magnitude as the
$\bar{s}$-quark density.
Samples of NC $\nu N$ and $\bar{\nu} N$ events containing a low
energy muon were also studied.
The muons in these events arise from the decay of hadrons produced in
the scattering.
Single muon production in NC $\nu N$ and $\bar{\nu} N$ DIS is
sensitive to the charm quark density.
The NuTeV results do not require the introduction of intrinsic charm.
CCFR\cite{Bib:Bodek} has determined the difference between the
structure function $xF_3^{\bar{\nu}}$ in CC $\bar{\nu} N$ scattering
and the structure function $xF_3^{\nu}$ in CC ${\nu} N$ scattering,
$\Delta({xF_3^\nu})$. 
In the quark parton model $\Delta({xF_3^\nu}) = 4 x (s - c)$.
$\Delta({xF_3^\nu})$ was presented in the kinematic range
$0.015<x<0.08$ and for $7<Q^2<20$~GeV$^2$.
The data are reasonably well described by NLO QCD using standard
PDF sets.

\subsection{The future of lepton-nucleon deep inelastic scattering}
\label{Sect:FutureDIS}

The upgrade to the HERA accelerator, which was nearing completion at
the time DIS01 was held, will yield a factor five more luminosity and
deliver polarised electron and positron beams to the collider
experiment ZEUS and H1.
The ZEUS and H1 experiments are also undergoing far reaching
upgrades. 
The upgrades themselves and the physics opportunities of the new
machine and detectors are described elsewhere in these proceedings
\cite{Bib:Elsen,Bib:Foster}.
Looking further ahead there are two new windows on lepton-nucleon deep
inelastic scattering that were discussed at DIS01.
The first, $ep$ collisions between 920~GeV protons from the HERA
proton ring and 250~GeV electrons or positrons from the proposed TESLA
linac, offers exciting possibilities in the study of low-$x$
DIS and diffraction\cite{Bib:Klein}.
The second, $\nu N$ scattering, would use the intense neutrino beam
generated by the decay in flight of the intense stored muon beam of a
future Neutrino Factory\cite{Bib:Morfin}.
The unprecedented neutrino flux that such a facility will provide
will allow neutrino targets with a mass of only a few kilogrammes 
to be used. 
Not only would experiments at such a facility be able to determine all
the unpolarised neutrino-nucleon structure functions with great
precision but the flux would be high enough that high precision
measurements using polarised nuclear targets will be possible. 
%
\section*{Acknowledgments}
 We very much enjoyed this well organised workshop in the beautiful town
 of Bologna, and we  thank the organisers for the support and
 hospitality given to us. We would also like to thank all those who
 contributed to the structure function session either by preparing
 talks or by taking part in the lively debates. We greatly appreciate
 the help that we have been given by the speakers in preparing the
 summary talks and in the preparation of these proceedings.
%
%

\end{document}